\RequirePackage[l2tabu, orthodox]{nag}
\documentclass[envcountsect,10pt]{llncs}
\usepackage[charter]{mathdesign}

\usepackage[T1]{fontenc}
\usepackage[utf8]{inputenc}
\usepackage{textcomp} % Euro and other unicode symbols
\PassOptionsToPackage{hyphens}{url}
\usepackage[bookmarksnumbered,bookmarksopen,breaklinks,colorlinks=true]{hyperref}
\usepackage[hyphenbreaks]{breakurl}
\usepackage{graphicx}
\usepackage{overpic}
\usepackage{amsmath}
\usepackage[english]{babel}
\usepackage[style=numeric,maxnames=5,giveninits=true,doi=false,isbn=false,url=false,eventdate=comp]{biblatex}
\usepackage[short]{bibstrings}
\usepackage{xspace}
\usepackage{verbatim}
\usepackage{general}
\usepackage{markup}
\usepackage{annot}
\usepackage{bodyversion}
\usepackage{environment}
\usepackage{hyphenation}

\AtEveryBibitem{%
\ifentrytype{inproceedings}{
  \clearname{editor}%
  \clearfield{booktitleaddon}%
}{}
}

\renewbibmacro{in:}{%
  \ifentrytype{article}{}{\printtext{\bibstring{in}\intitlepunct}}}
\DeclareFieldFormat[misc]{title}{\mkbibquote{#1}}
\DeclareSourcemap{
  \maps{
    \map{
      \step[typesource=tweet, typetarget=misc, final]
      \step[fieldset=howpublished, fieldvalue={Twitter}]
    }
    \map{
      \step[typesource=blog, typetarget=article, final]
      \step[fieldset=entrysubtype, fieldvalue={newspaper}]
      }
    \map{
      \step[typesource=movie, typetarget=misc, final]
      }
    \map{
      \step[typesource=newspaper, typetarget=article, final]
      \step[fieldset=entrysubtype, fieldvalue={newspaper}]
      }
    }
}
\renewcommand{\privkey}[1][{}]{{k_{#1}}}
\renewcommand{\pubkey}[1][{}]{{K_{#1}}}
\newcommand{\TEK}[1]{K_{#1}}
\newcommand{\EI}[2]{I_{#1,#2}}
\newcommand{\sentdb}{S}
\newcommand{\recdb}{R}
\let\mid=\|
\def\|{\,\mid\,}

\title{A Critique of the Google Apple Exposure Notification (GAEN) Framework\thanks{%
    Version: Tue Jan 12 10:08:06 2021 +0100 / arXiv2-1-g0210682 / gaen- critique.tex}
}
\author{Jaap-Henk Hoepman\inst{1,2}}
\institute{
  Radboud University Nijmegen, Email: \email{jhh@cs.ru.nl} \and
  University of Groningen
}
\begin{document}
\maketitle
\begin{abstract}
  As a response to the COVID-19 pandemic digital contact tracing has been proposed as a tool to support the health authorities in their quest to determine who has been in close and sustained contact with a person infected by the coronavirus. In April 2020 Google and Apple released the Google Apple Exposure Notification (GAEN) framework, as a decentralised and more privacy friendly platform for contact tracing. The GAEN framework implements exposure notification mostly at the operating system layer, instead of fully at the app(lication) layer.
  
  In this paper we study the consequences of this approach. We argue that this creates a dormant functionality for mass surveillance at the operating system layer. We show how it does not technically prevent the health authorities from implementing a purely centralised form of contact tracing (even though that is the stated aim). We highlight that GAEN allows Google and Apple to dictate how contact tracing is (or rather isn't) implemented in practice by health authorities, and how it introduces the risk of function creep.
\end{abstract}
\section{Introduction}

Large parts of the world are still suffering from a pandemic caused by the Severe Acute Respiratory Syndrome CoronaVirus 2 (SARS-CoV-2) that raised its ugly head somewhere late 2019, and that was first identified in December 2019 in Wuhan, mainland China~\autocite{vandorp2020emergence-sars-cov2}. Early work of Feretti~\etal~\autocite{ferretti2020covid}, modelling the infectiousness of SARS-CoV-2, showed that (under a number of strong assumptions) digital contact tracing could in principle help reduce the spread of the virus. This spurred the development of \term{contract tracing} apps,
(also known as \term{proximity tracing} or \term{exposure notification} apps),
that aim to support the health authorities in their quest to quickly determine who has been in close and sustained contact with a person infected by this virus~\autocite{who2020contact-tracing,martin2020demystifying-covid19-tracing}.

The main idea underlying digital contact tracing is that many people carry a smartphone most of the time, and that this smart phone could potentially be used to more or less automatically collect information about people someone has been in close contact with. Even though the effectiveness of contact tracing is contested~\autocite{bay2020no-panacea} and there are ethical concerns~\autocite{morley2020ethics-covid19}, especially Bluetooth-based contact tracing apps have quickly been embraced by governments across the globe (even though Bluetooth signal strength is a rather poor proxy for being in close contact~\autocite{dehaye2020bluetooth}). Bluetooth based contact tracing apps broadcast an ephemeral identifier on the short range Bluetooth radio network at regular intervals, while at the same time collecting such identifiers transmitted by other smartphones in the vicinity. The signal strength is used as an estimate for the distance between the two smartphones, and when this distance is determined to be short (within $1$--$2$ meters) for a certain period of time (typically $10$--$20$ minutes) the smartphones register each others ephemeral identifier as a potential risky contact.

Contact tracing is by its very nature a privacy invasive affair, but the level of privacy infringement depends very much on the particular system used. In particular it makes a huge difference whether \emph{all} contacts are registered \emph{centrally} (on the server of the national health authority for example) or in a \emph{decentralised} fashion (on the smartphones of the users that installed the contact tracing app)~\autocite{martin2020demystifying-covid19-tracing,hoepman2021hansel}.\footnote{%
  Note that essentially all systems for contact tracing require a central server to coordinate some of the tasks. The distinction between centralised and decentralised systems is therefore \emph{not} made based on whether such a central server exist, but based on where the matching of contacts takes place.
}
In the first case, the authorities have a complete and perhaps even real time view of the social graph of all participants. In the second case, information about one's contacts is only released (with consent) when someone tests positive for the virus.

One of the first contact tracing systems was the TraceTogether app deployed in Singapore.\footnote{%
  See \url{https://www.tracetogether.gov.sg}.
}
This inherently centralised approach lets phones exchange regularly changing pseudonyms over Bluetooth. A phone of a person who tests positive is requested to submit all pseudonyms it collected to the central server of the health authorities, who are able to recover phone numbers and identities from these pseudonyms. The Pan-European Privacy-Preserving Proximity Tracing (PEPP-PT) consortium quickly followed suit with a similar centralised proposal for contact tracing to be rolled out in Europe.\footnote{%
  See this WikiPedia entry \url{https://en.wikipedia.org/wiki/Pan-European_Privacy-Preserving_Proximity_Tracing} (the original domain \url{https://www.pepp-pt.org} has been abandoned, but some information remains on the project Github pages \url{https://github.com/pepp-pt}).
}
This consortium had quite some traction at the European policy level, but there were serious privacy concerns due to its centralised nature. As a response, a large group of academics, led by Carmela Troncoso and her team at EPFL, broke their initial engagement to the PEPP-PT consortium and
rushed to publish the Decentralized Privacy-Preserving Proximity Tracing (DP-3T) protocol\footnote{%
  See \url{https://github.com/DP-3T/documents}.
}
as a decentralised alternative for contact tracing with better privacy guarantees~\autocite{dp3t-whitepaper}. See~\autocite{veale2020sovereignty} for some details on the history.

All these protocols require low-level access to the Bluetooth network stack on a smartphone to transmit and receive the ephemeral identifiers used to detect nearby contacts. However, both Google's Android and Apple's iOS use a smartphone permission system to restrict access to critical or sensitive resources, like the Bluetooth network. This proved to be a major hurdle for practical deployment of these contact tracing apps, especially on iPhones as Apple refused to grant the necessary permissions.
%This reluctance is understandable given the fact that offering apps unfettered access to the Bluetooth network raises serious privacy concerns~\autocite{??}.
% \autocite{cunche2013mac-address} shows how a a MAC address can be linked to
% an actual person, and how this can subsequently be (ab)used to track people of interest. (But this is a different issue)

Perhaps in an attempt to prevent them from being manoeuvred into a position where they would have to grant to access to any contact tracing app (regardless of its privacy risk), Google and Apple instead developed their own platform for contact tracing called Google Apple Exposure Notification (GAEN). Around the time that DP-3T released their first specification to the public (early April 2020) Google and Apple released a joint specification for contact tracing as well (which they later updated and renamed to exposure notification\footnote{%
   See \url{https://techcrunch.com/2020/04/24/apple-and-google-update-joint-coronavirus-tracing-tech-to-improve-user-privacy-and-developer-flexibility/}.
})
with the aim to embed the core technology in the operating system layer of both recent Android and iOS powered smartphones\footnote{%
  See \url{https://www.google.com/covid19/exposurenotifications/} and
  \url{https://www.apple.com/covid19/contacttracing/}.
}.
Their explicit aim was to offer a more privacy friendly of exposure notification (as it is based on a distributed architecture) instead of allowing apps direct access to the Bluetooth stack to implement contact tracing or exposure notification themselves.

In this paper we study the consequences of pushing exposure notification down the stack from the app(lication) layer into the operating system layer. We first explain how contact tracing and exposure notification works when implemented at the app layer in section~\ref{sec-how}. We then describe the GAEN framework in section~\ref{sec-gaen}, and the technical difference between the two approaches in section~\ref{sec-difference}. Section~\ref{sec-critique} then discusses the concerns raised by pushing exposure notification down the stack: it creates a dormant functionality for mass surveillance at the operating system layer, it does not technically prevent the health authorities from implementing a purely centralised form of contact tracing (even though that is the stated aim), it allows Google and Apple to dictate how contact tracing is (or rather isn't) implemented in practice by health authorities, and it creates risks of function creep.\footnote{%
  This paper is based on two blog posts written by the author earlier this year: \url{https://blog.xot.nl/2020/04/19/google-apple-contact-tracing-gact-a-wolf-in-sheeps-clothes/} and 
  \url{https://blog.xot.nl/2020/04/11/stop-the-apple-and-google-contact-tracing-platform-or-be-ready-to-ditch-your-smartphone/}.
}

\section{How contact tracing and exposure notification works}
\label{sec-how}

%Contact tracing is one of the traditional tools used to fight an epidemic. The goal of contact tracing is to find all recent contacts of a patient that tested positive for infection of a virus. The idea being that these contacts could have been infected by this patient. By tracing these contacts, testing them, and treating or quarantining them, spread of the virus can be contained~\autocite{who2020contact-tracing}. Contact tracing is typically done 'by hand' by the national health authorities, and is a time consuming process. Digital contact tracing is supposed to support this traditional form of contact tracing, as people may not necessarily know or remember all people they have recently been in contact with.

As mentioned before, centralised systems for digital contact tracing automatically register all contacts of all people that installed the contact tracing app in a central database maintained by the health authorities. Once a patient tests positive, their contacts can immediately retrieved from this database. But as the central database is collecting contacts regardless of infection, someone's contacts can be retrieved by the authorities regardless. Hence the huge privacy risks associated with such a centralised approach.

Decentralised systems for digital contact tracing only record contact information locally on the smartphones of the people that installed the app: there is no central database. Therefore, the immediate privacy risk is mitigated.\footnote{%
  Certain privacy risks remain, however. See for example~\autocite{vaudenay2020dp3t,dp3t-whitepaper}.
}
Once a person tests positive however, some of the locally collected data is revealed. Some schemes (\eg DESIRE\footnote{%
  See
  \url{https://github.com/3rd-ways-for-EU-exposure-notification/project-DESIRE}.
},
and see~\autocite{hoepman2021hansel}) reveal the identities of the people that have been in close contact with the person that tested positive to the health authorities. Those variants are still contact tracing schemes. Most distributes schemes however only notify the persons that have been in close contact with the person that tested positive by displaying a message on their smartphone. The central health authorities are not automatically notified (and remain in the dark unless the people notified take action and get tested, for example). Such systems implement \emph{exposure notification}.

\begin{figure}
  \centering
  \includegraphics{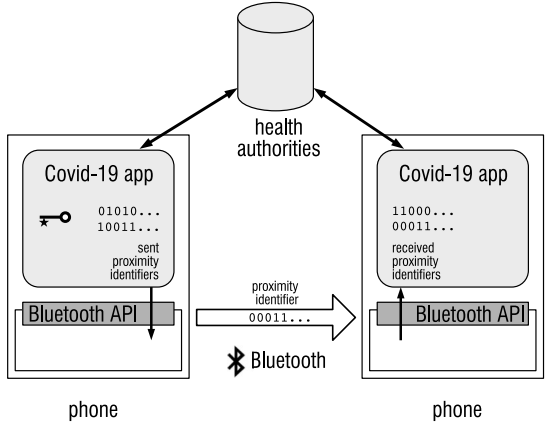}
  \caption{Exposure notification using an app}
  \label{fig-en}
\end{figure}

Most exposure notification systems (like DP-3T and GAEN) distinguish a  \emph{collection} phase and a \emph{notification} phase, that each work roughly as follows (see also figure~\ref{fig-en}). 

During the collection phase the smartphone of a participating user generates a random ephemeral proximity identifier $\EI{d}{i}$ every $10$--$20$ minutes, and broadcasts this identifier over the Bluetooth network every few minutes.\footnote{
  In this notation $\EI{d}{i}$ denotes the ephemeral proximity identifier generated for the $i$-th $10$--$20$ minute time interval on day $d$.
}
The phone also stores a copy of this identifier locally. The smartphones of other nearby participating users receive this identifier and, when the signal strength indicates the user is within the threshold distance of $1$--$2$ meters, stores this identifier (provided it sees the same identifier several times within a $10$--$20$ minute time interval). A smartphone of a participant thus contain a database $\sentdb$ of identifiers it sent itself and another database $\recdb$ of identifiers it received from others. The time an identifier was sent or received is also stored, at varying levels of precision (in hours, or days, for example). The databases are automatically pruned to delete any identifiers that are no longer epidemiologically relevant (for COVID-19, this is any identifier that was collected more than $14$ days ago).

The notification phase kicks in as soon as a participating user tests positive for the virus and agrees to notify their contacts. In that case the user instructs their app to upload the database $\sentdb$ of identifiers the app sent itself to the server of the health authorities.\footnote{%
  A contact tracing version of the app would instead request the smartphone of the user to upload the database $\recdb{}$ of \emph{received} identifiers that the app collected.
}
The smartphone app of other participants regularly queries this server for recently uploaded identifiers of infected people, and matches any new identifiers it receives from the server with entries in the database of identifiers it received from others in close proximity the last few weeks. If there is a match, sometime during the last few weeks the app must have received and stored an identifier of someone who just tested positive for the virus. The app therefore notifies its user that they have been in close contact with an infected person recently (sometimes indicating the day this contact took place). It then typically offers advice on how to proceed, like offering pointers to more information, and strongly suggesting to contact the health authorities, get tested, and go into self-quarantine.

\section{The GAEN framework}
\label{sec-gaen}

Google and Apple's framework for exposure notification follows the same paradigm, with the notable exception that instead of an app implementing all the functionality, most of the framework is implemented at the operating system layer instead.\footnote{%
  The summary of GAEN and its properties is based on the documentation offered by both Google (\url{https://www.google.com/covid19/exposurenotifications/}) and Apple
  (\url{https://www.apple.com/covid19/contacttracing/}) online, and was last checked December 2020.
  The documentation offered by Google and Apple is terse and scattered. 
  The extensive documentation of the Dutch CoronaMelder at
  \url{https://github.com/minvws} proved to be very helpful.
}
Although GAEN is a joint framework, there are minor differences in how it is implemented on Android (Google) and iOS (Apple).
GAEN works on Android version 6.0 (API level 23) or higher, and on some devices as low as version 5.0 (API level 21).\footnote{%
  See \url{https://developers.google.com/android/exposure-notifications/exposure-notifications-api}.
}
On Android GAEN is implemented as a Google Play service. GAEN works for Apple devices running iOS 13.5 or higher.

At the Bluetooth and 'cryptographic' layer GAEN works the same on both platforms however. This implies that ephemeral proximity identifiers sent by any Android device can be received and interpreted by any iOS device in the world and vice versa. In other words: users can \emph{in principle} get notified of exposures to infected people independent of the particular operating system their smartphone runs, and independent of which country they are from. (In practice some coordination between the exposure notification apps and the back-end servers of the different health authorities involved is required.)

\begin{figure}
  \centering
%  \setoverpicfontsize
  \begin{overpic}[abs,unit=1pt]{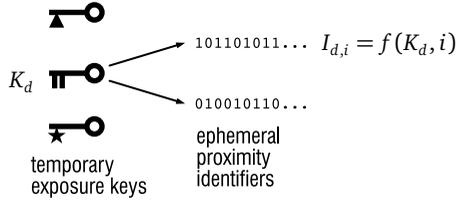}%
    \def\k{$\TEK{d}$}
    \def\i{$\EI{d}{i} = f(\TEK{d},i)$}
  \input{./fig/gaen-keys.overpic}%
  \end{overpic}
  \caption{Temporary exposure keys and ephemeral proximity identifiers.}
  \label{fig-gaen-keys}
\end{figure}

As an optimisation step, devices do not randomly generate each and every ephemeral proximity identifier independently. Instead, the ephemeral proximity identifier $\EI{d}{i}$ to use for a particular interval $i$ on day $d$ is derived from a \emph{temporary exposure key} $\TEK{d}$ (which \emph{is} randomly generated each day) using some public deterministic function $f$ (the details of which do not matter for the current paper). In other words $\EI{d}{i} = f(\TEK{d},i)$, see figure~\ref{fig-gaen-keys}. With this optimisation, devices only need to store exposure keys in $\sentdb$, as the actual ephemeral proximity identifiers can always be reconstructed from these keys.

\begin{figure}
  \centering
  \includegraphics{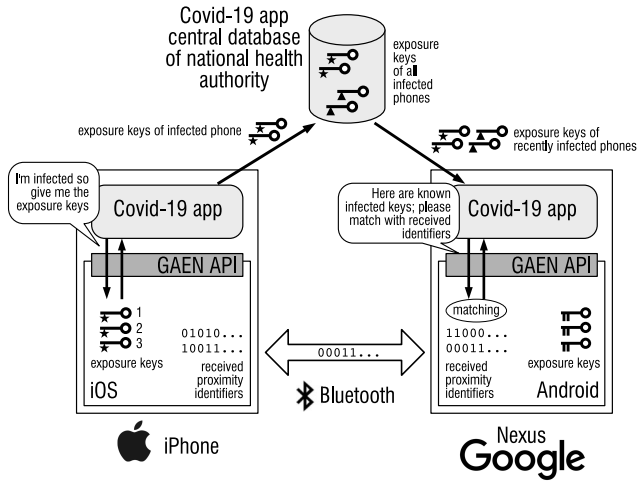}
  \caption{The GAEN framework.}
  \label{fig-gaen}
\end{figure}

Generating, broadcasting, and collecting ephemeral proximity identifiers happens automatically at the operating system layer, but only if the user has explicitly enabled this by installing an exposure notification app and setting the necessary permissions,\footnote{%
  Both Android and iOS require Bluetooth to be enabled. On Android 10 and lower, the device location setting needs to be turned on as well (see \url{https://support.google.com/android/answer/9930236}). 
}
or by enabling exposure notifications in the operating system settings.\footnote{%
  For those countries where the national health authorities have not developed their own exposure notification app but instead rely on Exposure Notification Express (see \url{https://developers.google.com/android/exposure-notifications/en-express}).
}
Apple and Google do not allow exposure notification apps to access your device location.\footnote{%
  See \url{https://support.google.com/android/answer/9930236} and \url{https://covid19-static.cdn-apple.com/applications/covid19/current/static/contact-tracing/pdf/ExposureNotification-FAQv1.2.pdf}.
}
By default, exposure notification is disabled on both platforms. When enabled, the database of $\sentdb$ of exposure keys and the database $\recdb$ of identifiers received are stored at the operating system layer, which ensures that data is not directly accessible by any app installed by the user.

Actual notifications are the responsibility of the exposure notification app. In order to use the data collected at the operating system layer, the app needs to invoke the services of the operating system through the GAEN Application Programming Interface (API). Apps can only access this API after obtaining explicit permission from Google or Apple. The API offers the following main functions (see also figure~\ref{fig-gaen}).
\begin{itemize}
\item
  \emph{Retrieve} the set of exposure keys (stored in $\sentdb$). ``The app must provide functionality that confirms that the user has been positively diagnosed with COVID-19.''\footnote{%
  See \url{https://developers.google.com/android/exposure-notifications/exposure-notifications-api}. Also see the verification system Google designed for this
  \url{https://developers.google.com/android/exposure-notifications/verification-system}.
  }
  % https://developer.apple.com/documentation/exposurenotification/supporting_exposure_notifications_express
  But this is not enforced at the API layer. In other words, the app (once approved and given access to the API) has access to the exposure keys.
\item
  \emph{Match} a (potentially large) set of exposure keys against the set of ephemeral proximity identifiers received from other devices earlier (stored in $\recdb$), and return a list of risk score (either a list of daily summaries, or a list  of individual $<30$ minute exposure windows). This function is rate limited to a few calls per day.\footnote{%
     On iOS 13.7 and the most recent version of the API limits the use of this method to a maximum of six times per 24-hour period. On Android, the most recent version of the API also allows at most six calls per day, but 'allowlisted' accounts (used by health authorities for testing purposes) are allowed 1,000,000 calls per day).
  }
\end{itemize}
The API also ensures that the user is asked for consent whenever an app enables exposure notification for the first time, and whenever user keys are retrieved for upload to the server of the health authorities after the user tested positive for COVID-19. The API furthermore offers functions to tune the computation of the risk scores.

The idea is that through the API a user that test positive for COVID-19 can instruct the app to upload all its recent (the last 14, actually) temporary exposure keys to the server of the health authorities. The exposure notification app of another user can regularly query the server of the health authorities for recently uploaded exposure keys of infected devices. Using the second GAEN API function allows the app to submit these exposure keys to the operating system which, based on the database $\recdb$ of recently collected proximity identifiers, checks whether there is a match with such an exposure key (by deriving the proximity identifiers locally). A list of matches is returned that contains the day of the match and an associated risk score; the actual key and identifier matched is not returned however. The day of the contact, the duration of the contact, the signal strength (as a proxy for distance of contact), and the type of test used to determine infection are used to compute the risk score. Developers can influence on how this risk score is computed by providing weights for all the parameters. Using the returned list the app can decide to notify the user when there appears to be a significant risk of infection. Note that by somewhat restricting the way risk scores are computed, GAEN makes it harder for a malicious app to determine exactly which exposure key triggered a warning (and hence makes it harder to determine exactly with whom someone has been physical proximity with).

\section{How the GAEN framework differs from a purely app based approach}
\label{sec-difference}

Given its technical architecture, the GAEN framework fundamentally differs from an purely app based approach to exposure notification in the following four aspects.

First of all, the functionality and necessary code for the core steps of exposure notification (namely broadcasting, collecting and matching ephemeral proximity identifiers) comes pre-installed on all modern Google and Apple devices. In a purely app based approach this functionality and code is solely contained in the app itself, and not present on the device when the app is not installed (and removed when the app is de-installed).

Second, all relevant data (ephemeral proximity identifiers and their associated metadata like date, time and possibly location) are collected and stored at the operating system level. In a purely app based approach this data is collected and stored at the user/app level. This distinction is relevant as in modern computing devices the operating system runs in a privileged mode that renders the data it processes inaccessible to 'user land' app. Data processed by apps is accessible to the operating system in raw form (in the sense the operating system has access to all bytes of memory used by the app), but the interpretation of that data (which information is stored where) is not necessarily easy to determine. 

Moreover, the framework is interoperable at the global level: users can \emph{in principle} get notified of exposures to infected people independent of the particular operating system their smartphone runs, and independent of which country they are from. This would not necessarily be the case (and probably in practice be impossible to achieve) in a purely app based approach.

Finally the modes of operation are set by Google and Apple: the system notifies users of exposure; it does not automatically inform the health authorities. The app is limited to computing a risk score, and does not receive the exact location nor the exact time when a 'risky' contact took place. In a purely app based approach the developers of the app themselves determine the full functionality of the app (within the possible limits imposed by the app stores).

\section{A critique of the GAEN framework}
\label{sec-critique}

It is exactly for the above properties that the GAEN framework appears to protect privacy: the health authorities (and the users) are prevented from obtaining details about the time and location of a risky contact, thus protecting the privacy of infected individuals, and the matching is forced to take place in a decentralised fashion (which prevents the health authorities from directly obtaining the social graph of users).

However, there is more to this than meets the eye, and there are certainly broader concerns that stem from the way GAEN works.
\begin{itemize}
\item
  By pushing exposure notification down the stack, GAEN creates a dormant functionality for mass surveillance at the operating system layer.
\item
  Moreover, the exposure notification microdata (exposure keys and proximity identifiers) are under Google/Apple's control.
\item
  GAEN does not \emph{technically prevent} health authorities from implementing a purely centralised form of contact tracing (although it clearly discourages it). A decentralised framework like GAEN can be re-centralised. The actual protection offered therefore remains \emph{procedural}: we need to trust Google and Apple to disallow centralised implementations of contact tracing apps offered through their app stores.
\item
  GAEN leverages Google and Apple's control over how exposure notification works because exposure notification apps are required to use it. In particular it allows Google and Apple to dictate how contact tracing is (or rather isn't) implemented in practice by health authorities.
\item
  GAEN introduces significant risks of function creep.
\end{itemize}
These concerns are discussed in detail in the following sections. These are by no means the only ones,\footnote{%
  See \url{https://www.eff.org/deeplinks/2020/04/apple-and-googles-covid-19-exposure-notification-api-questions-and-answers}.
}
see for example also~\eg\autocite{sharon2020blind-sided,duarte2020gaen,klein2020corona}, but these are the ones that derive directly from the architectural choices made.

%For more information we refer to ``Data Protection Law \& COVID-19: An Observatory'' maintained by VUB's Law, Society, Technology \& Society research group,\footnote{%
%  \url{https://lsts.research.vub.be/en/data-protection-law-and-the-covid-19-outbreak}.
%}
%and to the ``Global Perspectives'' book~\autocite{taylor2020data-justice-covid19} published by the Global Data Justice project.\footnote{%
%  \url{https://globaldatajustice.org}.
%}

\subsection{GAEN creates a dormant mass surveillance tool}

Instead of implementing all exposure notification functionality in an app, Google and Apple push the technology down the stack into the operating system layer creating a Bluetooth-based exposure notification platform. This means the technology is available all the time, for all kinds of applications beyond just exposure notification. As will explained in the next section, GAEN can be (ab)used to implement centralised forms of contact tracing as well. Exposure notification is therefore no longer limited in time, or limited in use purely to trace and contain the spread of COVID-19. This means that two very important safeguards to protect our privacy are thrown out of the window.

Moving exposure notification down the stack fundamentally changes the amount of control users have: you can uninstall a (exposure notification) app, you cannot uninstall the entire OS (although on Android you can in theory disable and even delete Google Play Services). The only thing a user can do is disable exposure notification using an operating system level setting. But this does not remove the actual code implementing this functionality.

But the bigger picture is this: it creates a platform for contact tracing in the more general sense of mapping who has been in close physical contact (regardless of whether there is a pandemic that needs to be fought).  Moreover,this platform for contact tracing works all across the globe for most modern smart phones (Android Marshmallow and up, and iOS 13 capable devices) across both OS platforms. Unless appropriate safeguards are in place this would create a global mass-surveillance system that would reliably track who has been in contact with whom, at what time and for how long.\footnote{%
  GAEN does not currently make all this information available in exact detail through its API, but it \emph{does} collect this information at the lower operating system level. It is unclear whether GAEN records location data at all (although it would be easy to add this, and earlier versions of the API did in fact offer this information). 
}
GAEN works much more reliably and extensively to determine actual physical contact than any other system based on either GPS or mobile phone location data (based on cell towers) would be able to (under normal conditions). It is important to stress this point because some people believe this is something companies like Google (using their GPS and WiFi names based location history tool) can already do for years. This is not the case. This type of contact tracing really brings it to another level.

In those regions that that opt for Exposure Notification Express\footnote{%
  See \url{https://developers.google.com/android/exposure-notifications/en-express}.
}
the data collection related to exposure notification starts as soon as you accept the operating system update and enable it in the settings. In other regions this only happens when people install a exposure notification app that uses the API to find contacts based on the data phones have already collected. But this assumes that both Apple and Google indeed refrain from offering other apps access to the exposure notification platform (through the API) through force or economic incentives, or suddenly decide to use the platform themselves. GAEN creates a dormant functionality for mass surveillance~\autocite{veale2020gact}, that can be turned on with the flip of a virtual switch at Apple or Google HQ.

All in all this means we all have to put a massive trust in Apple and Google to properly monitor the use of the GAEN API by others.

\subsection{Google and Apple control the exposure notification microdata}

Because the exposure notification is implemented at the operating system layer, Google and Apple fully control how it works and have full access to all microdata generated and collected. In particular they have, in theory, full access to the temporary exposure keys and the ephemeral proximity identifiers, and control how these keys are generated. We have to trust that the temporary exposure keys are really generated at random, and not stealthily derived from a user identifier that would allow Google or Apple to link proximity identifiers to a particular user. And even if these keys are generated truly at random, at any point in time Google or Apple could decide to surreptitiously retrieve these keys from certain device, again with the aim to link previously collected proximity identifiers to this particular device. In other words, we have to trust Google and Apple will not abuse GAEN themselves. They do not necessarily have an impeccable track record that warrants such trust.

\subsection{Distributed can be made centralised}

The discussion in the preceding paragraphs implicitly assumes that the GAEN platform truly enforces a decentralised form of exposure notification, and that it prevents exposure notification apps from automatically collecting information on a central server about who was in contact with who. This assumption is not necessarily valid however (although it can be enforced provided Apple and Google are very strict in the vetting process used to grant apps access to the GAEN platform). In fact, GAEN can easily be used to create a centralised from of exposure notification, at least when we limit our discussion to centrally storing information about who has been in contact with an infected person.

The idea is as follows. GAEN allows a exposure notification app on a phone to test daily exposure keys of infected users against the proximity identifiers collected by the phone over the last few days. This test is local; this is why GAEN is considered decentralised. However, the app could immediately report back the result of this test to the central server, without user intervention (or without the user even noticing).\footnote{%
  Recall that the API enforces user consent when retrieving exposure keys, but not when matching them.
}
It could even send a user specific identifier along with the result, thus allowing the authorities to immediately contact anybody who has recently been in the proximity of an infected person. This is the hallmark of a centralised solution.

In other words: the GAEN technology itself does not prevent a centralised solution. The only thing preventing it would be Apple and Google being strict in vetting exposure notification apps. But they could already do so now, without rolling out their GAEN platform, by strictly policing which apps they allow access to the Bluetooth networks stack, and which apps they allow on their app stores.

A malicious app could do other things as well. By design GAEN does not reveal which infected person a user has been in contact with when matching keys on the users phone. Calls to the matching function in the API are rate limited to a few calls each day, the idea being that a large number of keys can be matched in batch without revealing which particular key resulted in a match. But this still allows a malicious app (and accompanying malicious server) to test a few daily tracing keys (for example of persons of interest) one by one, and to keep track of each daily tracing key for which the test was positive, and report these back to the server. As the server knows which daily tracing key belongs to which infected person, this allows the server to know exactly with which infected persons of interest the user of this phone has been in contact with. If the app is malicious, even non-infected persons are at risk, because such an app could retrieve the exposure notification keys even if a user is not infected (provided it can trick the user in consenting to this).

Clearly a malicious exposure notification app not based on GAEN could do the same (and much more). But this does shows that GAEN by itself does not protect against such scenarios, while making the impact of such scenarios far greater because of its global reach.

\subsection{Google and Apple dictate how contact tracing works}

Apple and Google’s move is significant for another reason: especially on Apple iOS devices, access to the hardware is severely restricted. This is also the case for access to Bluetooth. In fact, without approval from Apple, you cannot use Bluetooth ‘in the background’ for your app (which is functionality that you need to collect information about nearby phones even if the user phone is locked). You could argue that this could potentially \emph{improve} privacy as it adds another checkpoint where some entity (in this case Apple) decides whether to allow the proposed app or not. But Apple (and by extension Google) use this power as a leverage to grab control over how contact tracing or exposure notification will work. This is problematic as this allows them to set the terms and conditions, without any form of oversight. With this move, Apple and Google make themselves indispensable, ensuring that this potentially global surveillance technology is forced upon us. And as a consequence all microdata underlying any contact tracing system is stored on the phones they control.

For example, the GAEN framework prevents notified contacts to learn the nature of the contact and make well informed decision about the most effective response: get tested, or go into self-quarantine immediately. It also prevents the health authorities from learning the nature of the contact and hence makes it impossible to build a model of contacts. ``The absence of transmission data limits the scope of analysis, which might, in the future, give freedom to people who can work, travel and socialise, while more precisely targeting others who risk spreading the virus.''~\autocite{ilves2020google-apple-dictating}. This happens because the GAEN framework is based on a rather corporate understanding of privacy as giving control and by asking for consent. But under certain specific conditions, and a public health emergency like the current pandemic is surely one, individual consent is not appropriate: ``to be effective, public-health surveillance needs to be comprehensive, not opt-in.''~\autocite{cohen2020danger-tech-covid19}.

%Repeats the same arguments
%\Todo{https://www.nytimes.com/2020/06/16/world/europe/contact-tracing-apps-europe-coronavirus.html?referringSource=articleShare}

\subsection{Function creep}

The use of exposure notification functionality as offered through GAEN is not limited to controlling just the spread of the COVID-19 virus. As this is not the first corona type virus, it is only a matter of time until a new dangerous virus will roar its ugly head. In other words, exposure notification is here to stay.

And with that, the risk of function creep appears: with the technology rolled out and ready to be (re)activated, other uses of exposure notification will at some point in time be considered, and deemed proportionate. Unless Apple and Google strictly police the access to the GAEN API (based on some publicly agreed upon rules) and ensure that it is only used by the health authorities, and only for controlling a pandemic like COVID-19, the following risks are apparent.

Consider the following hypothetical example of a government that wants to trace the contact or whereabouts of certain people, that could ensue when Google and Apple fail to strictly enforce access. Such a government could coerce developers to embed this tracking technology in innocent looking apps, in apps that you are more or less required to have installed, or in software libraries used by such apps. Perhaps it could even coerce Apple and Google itself to silently enable exposure notifications for all devices sold in their country, even if the users do not install any app.\footnote{%
  Note that when Google and Apple first announced their exposure notification platform the idea was that your phone would start emitting and collecting proximity identifiers as soon as the feature was enabled in the operating settings, even if no exposure notification app was installed, see~\url{https://techcrunch.com/2020/04/13/apple-google-coronavirus-tracing/}.
}
It is known that Google and Apple in some cases do bow to government pressure to enable or disable certain features: like filter search results\footnote{%
  See \url{https://www.sfgate.com/business/article/Google-bows-to-China-pressure-2505943.php}.
},
remove apps from the app store\footnote{%
  See \url{https://www.nytimes.com/2019/10/10/business/dealbook/apple-china-nba.html}.
}
and even move cloud storage servers,\footnote{%
  See \url{https://www.reuters.com/article/us-china-apple-icloud-insight/apple-moves-to-store-icloud-keys-in-china-raising-human-rights-fears-idUSKCN1G8060}.
}
offering Chinese authorities far easier access to text messages, email and other data stored in the cloud.

Because the ephemeral proximity identifiers are essentially random they cannot be authenticated. In other words: any identifier with the right format advertised on the Bluetooth with the correct service identifier will be accepted and recorded by any device with GAEN active. Moreover, because the way ephemeral identifiers are generated from daily exposure keys is (necessarily) public, anybody can build a cheap device broadcasting ephemeral identifiers from chosen daily exposure keys that will be accepted and stored by a nearby device with the GAEN platform enabled. A government could install such Bluetooth beacons at fixed locations of interest for monitoring purposes. The daily exposure keys of these devices could be tested against phones of people of interest running the apps as explained above. Clearly this works only for a limited number of locations because of rate limiting, but note that at least under Android this limit is not imposed for `allowlisted' apps for testing purposes, and then the question is again whether Google can be forced to allowlist a certain government app. China could consider it to further monitor Uyghurs. Israel could use it to further monitor Palestinians. You could monitor the visitors of abortion clinics, coffee shops, gay bars, \ldots 
Indeed the exact same functionality offered by exposure notification could allow the police to quickly see who has been close to a murder victim: simply report the victims phone as being 'infected'. Some might say this is not a bug but a feature, but the same mechanism could be used to find whistleblowers, or the sources of a journalist.

For centralised contact tracing apps we already see function creep creeping in.
The recent use of the term `contact tracing' in the context of tracking protesters in Minnesota after demonstrations erupted over the death of George Floyd at the hands of a police officer\footnote{%
  See  \url{https://bgr.com/2020/05/30/minnesota-protest-contact-tracing-used-to-track-demonstrators/}.
}
is ominous, even if the term refers to traditional police investigating methods\footnote{%
  And what to think of the following message posted by Anita Hazenberg, the Director Innovation Directorate at Interpol: ``Is your police organisation considering how tracing apps will influence the way we will police in the future? If you are a (senior) officer dealing with policy challenges in this area, please join our discussion on Wednesday 6 May (18.00 Singapore time) during a INTERPOL Virtual Discussion Room (VDR). Please contact EDGCI-IC@INTERPOL.INT for more info. Only reactions from law enforcement officers are appreciated.''. See:
\url{https://www.linkedin.com/posts/anita-hazenberg-b0b48516_is-your-police-organisation-considering-how-activity-6663040380965130242-q8Vk}.
}.
More concrete evidence is the discovery that Australia’s intelligence agencies were `incidentally' collecting data from the country’s COVIDSafe contact-tracing app.
\footnote{%
  \url{https://techcrunch.com/2020/11/24/australia-spy-agencies-covid-19-app-data/}
}
The Singapore authorities recently announced that the police can access COVID-19 contact tracing data for criminal investigations.\footnote{%
  \url{https://www.zdnet.com/article/singapore-police-can-access-covid-19-contact-tracing-data-for-criminal-investigations/}
}
Now one could argue that these examples are an argument supporting the privacy friendly approach taken by Google and Apple. After all, by design exposure notification does not have a central database that is easily accessible by law enforcement or intelligence agencies. But as explained above, this is not (and cannot be) strictly enforced by the GAEN framework. 

Contact tracing also has tremendous commercial value. A company could install Bluetooth beacons equipped with this software at locations of interest (e.g. shopping malls). By reporting a particular beacon as 'infected' all phones (that have been lured into installing a loyalty app or that somehow have the SDK of the company embedded in some of the apps they use) will report that they were in the area. Facebook used a crude version of contact tracing (using the access it had to WhatsApp address books) to recommend friends on Facebook~\autocite{tait2019facebook-friends,hill2016facebook-psychiatrist}. The kind of contact tracing offered by GAEN (and other Bluetooth based systems) gives a much more detailed, real time, insight in people’s social graph and its dynamics. How much more precise could targeted adverting become? Will Google and Apple forever be able to resist this temptation? If you have Google Home at home, Google could use this mechanism to identify all people that have visited your place. Remember: they set the restrictions on the API. They can at any time decide to change and loosen these restrictions. 

% Jealous partners could secretly install an app on the phone of their significant other, to allow them to monitor who they have been in contact with. Overzealous parents could use this spy on their children.

\section{Conclusion}
\label{sec-conclusion}

We have described how the shift, by Google and Apple, to push exposure notification down the stack from the app layer to the operating system layer fundamentally changes the risk associated with exposure notification systems, and despite the original intention, unfortunately not for the better. We have shown that from a technical perspective it creates a dormant functionality for global mass surveillance at the operating system layer, that it takes away the power to decide how contact tracing works from the national health authorities and the national governments, and how it increases the risks of function creep already nascent in digital exposure notification and contact tracing systems.
These risks can only be mitigated by Google and Apple as they are the sole purveyors of the framework and have sole discretionary power over who to allow access to the framework, and under which conditions. We fully rely on their faithfulness and vigilance to enforce the rules and restrictions they have committed to uphold, and have very little tools to verify this independently.

%
% use Makefile.main and friends to extract bib entries from main bib files
% and create a local .bib file; add its name here
%
% strings do not need to be included here (better portability); they are 
% merged into the local bibfile by Makefile.jhh
%
\setlength{\emergencystretch}{8em}
\printbibliography
\end{document}